\PassOptionsToPackage{colorlinks}{hyperref}
\documentclass{elsart}
\pdfoutput=1
\usepackage[pdftex]{graphicx}
\usepackage[unicode=true,
 bookmarks=true,bookmarksnumbered=true,bookmarksopen=false, breaklinks=false,backref=false,pagebackref=false, colorlinks=true]{hyperref}
\hypersetup{
pdftitle={Iterative method to compute the Fermat points and Fermat distances
of multiquarks},
pdfauthor={P. Bicudo, M. Cardoso},linkcolor=black, citecolor=black,  urlcolor=black,  filecolor=black, pdfpagelayout=OneColumn,pdfnewwindow=true,
pdfstartview=XYZ, plainpages=false}

\usepackage{amssymb}
\usepackage{amsmath}
\usepackage{subfig}
\usepackage{float}
\usepackage{url}
\usepackage[T1]{fontenc}
\usepackage[latin1]{inputenc}
\usepackage[english]{babel}

\newcommand{\be}{\begin{equation}}
\newcommand{\bea}{\begin{eqnarray}}
\newcommand{\ee}{\end{equation}}
\newcommand{\eea}{\end{eqnarray}}

\newcommand{\mbf}{\mathbf}

\begin{document}
\begin{frontmatter}
\title{Iterative method to compute the Fermat points and Fermat distances
of multiquarks}
\author{P. Bicudo and M. Cardoso}
\address{CFTP, Departamento de Física, Instituto Superior Técnico, Av. Rovisco Pais, 1049-001 Lisboa, Portugal}

\begin{abstract}
The multiquark confining potential is proportional to the total distance of the 
fundamental strings linking the quarks and antiquarks. We address the computation of the total string distance an of the Fermat points where the
different strings meet. For a meson (quark-antiquark system)
the distance is trivially the quark-antiquark distance. 
For a baryon (three quark system) the problem was solved geometrically
from the onset, by Fermat and by Torricelli. The geometrical
solution can be determined just with a rule and a compass, but
translation of the geometrical solution to an analytical expression is not 
as trivial.
For tetraquarks, pentaquarks, hexaquarks, etc, the geometrical solution is 
much more complicated.
Here we provide an iterative method, converging fast to the correct Fermat points 
and the total distances, relevant for the multiquark potentials. We also review briefly the geometrical methods leading to the Fermat points and to the total distances. 
\end{abstract}
\end{frontmatter}


\section{Introduction}

Fermat proposed to Torricelli the problem of finding the point in a triangle minimizing the sum of the distances to the three respective vertices. 
This first Fermat point or Torricelli point
\cite{Weisstein,Gallatly,Greenberg,Lindt,Johnson,Kimberling1,Kimberling2,Mowaffaq,Spain}, 
is the isogonic point,  since in a sufficiently acute triangle the angle formed 
by the segments connecting any two vertices with it is 120 degrees. 

Lately this problem became relevant 
for quark physics because the multiquark confining potential is proportional to 
the total distance of the fundamental strings linking the quarks and antiquarks.
Of course, we address here the case where we have a single multiquark and
not many free or molecular mesons and baryons, where the confining potential would be different.
The three-body star-like potential has already been used long ago
in Baryons
\cite{Carlson:1982xi}, however for many years there was a debate
in the lattice QCD community on the two-body versus three body
nature of the confining potential for baryons.
Recently, the study  of 
flux tubes  in Lattice QCD for Baryons (triquarks) by
Takahashi {\em et al} 
\cite{Takahashi:2000te}
confirmed the  three-body star-like confining potential .
Very recently, the Wilson loop technique
was applied to tetraquarks 
\cite{Okiharu:2004ve}
and pentaquarks
\cite{Okiharu:2004wy}
by Okiharu {\em et al},
showing that the confining potential is provided by a fundamental
string linking all the quarks and antiquarks. 
Cardoso {\em et al} also confirmed this result with the Wilson loops
for hybrids 
\cite{Bicudo:2007xp}
and for three gluon glueballs
\cite{Cardoso:2008sb}. 
Thus we assume that the confining component of the multiquark potential is,
\bea
V_c(\mbf r_i) &=&  \sigma \  \sum _{i,a} r_{i\,a}\ , 
 \label{potential}
\\ \nonumber 
\mbf  r_{i\,a}&=&  \mbf r_i -\mbf r_a  \ ,
\eea
where $\sigma$ is the string tension, $r_i$ is the position of the quark or antiquark
 $Q_i$ ,  $r_a$ is the position of the Fermat point $F_a$, and we use respectively arab digits $i=1,2,3 \cdots$ for the quarks (antiquarks) 
and roman digits $a=I,II,III \cdots$ for the Fermat points. 
 Thus the Fermat problem of finding the paths minimizing the total 
distance is equivalent to
 the physics problem of computing the multiquark potential. 
Notice that  there are already some proposed experimental signals of tetraquarks, and the next generation of Hadronic Detectors may eventually observe multiquark hadrons.

\begin{figure}[t!]
\begin{centering}
\includegraphics[width=15cm]{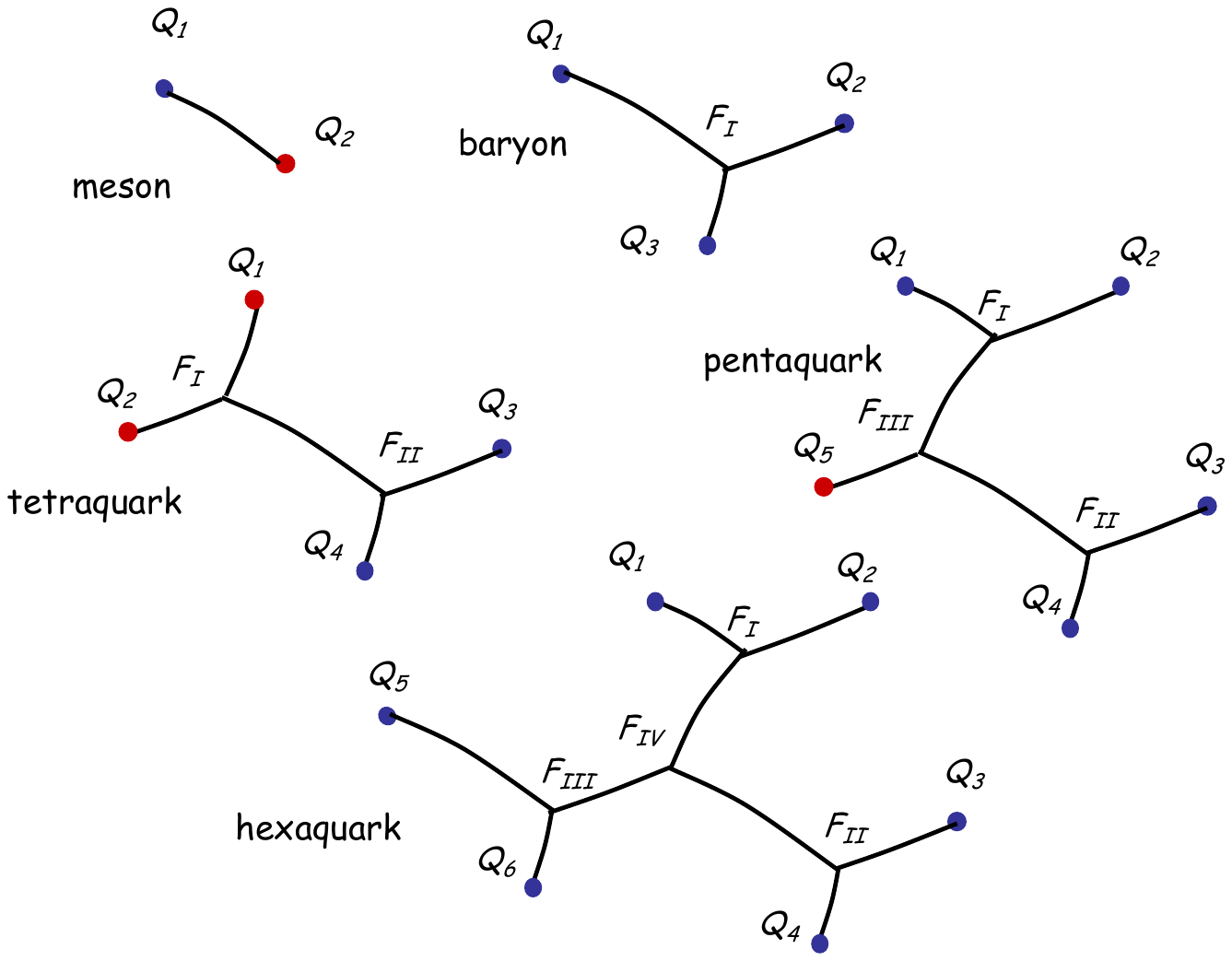}
\par\end{centering}
\caption{The geometries of the string sections linking the first five
multiquarks. Notice that the number of Fermat points $F_a$ is $N-2$ where $N$ is the number of quarks and antiquarks $Q_i$ in the multiquark. }
\label{first five}
\end{figure}

The geometries 
of the strings of the first five multiquarks are depicted in Fig. \ref{first five}.
Eq. (\ref{potential}) and Fig. \ref{first five}
extend the definition of the Fermat point of a triangle 
to the Fermat point of polygons in three dimensions with more points. 
With the present definition, where confinement is
produced by fundamental strings, the strings meet in internal three-string vertices.
The number of quarks can always be increased replacing a quark  (antiquark)
 by a Fermat point and a diquark (di-antiquark). Thus the number of quarks 
 (and antiquarks) minus the number of Fermat points is a constant. Since in the 
meson and baryon this constant is 2, the number of Fermat points is $N-2$ where 
$N$ is the number of quarks and antiquarks.  Moreover in eq. (\ref{potential}) 
we are only suming over distances between points linked by strings.

For a meson (quark-antiquark system)
the distance is trivially the quark-antiquark distance. 
For a baryon (three quark system) the problem was first solved geometrically by Fermat and by Torricelli. 
In the case of 3 quarks, the minimization of the potential in eq. (\ref{potential})
implies that,
\be
\widehat r_{1 \, I} + \widehat r_{2 \, I}+  \widehat r_{3 \, I}= 0 \ ,
\ee
and it is clear that the solution is that, either the triangle is not sufficiently acute,
or the angles are all equal to $120^o$,
\be
\widehat {\mbf r_{1 \, I},  \mbf r_{2 \, I}}
=
\widehat {\mbf r_{2 \, I},  \mbf r_{3 \, I}}
=  
\widehat {\mbf r_{3 \, I},  \mbf r_{1 \, I}}
= 120^o \ .
\label{isogonic}
 \ee
Due to the beauty of the triangles, and also to their simplicity, there are numerous geometry textbooks and articles on the Fermat - Torricelli point
\cite{Weisstein,Gallatly,Greenberg,Lindt,Johnson,Kimberling1,Kimberling2,Mowaffaq,Spain}. 
However, when the number of quarks increase, to  tetraquarks, pentaquarks, etc, the geometric construction of the Fermat points becomes more and more difficult. 
Thus a numerical solution of this problem is welcome.

Here we address the computation of the total string distance and of the Fermat points where the different strings meet.  
In Section 2 we review briefly the geometrical methods leading to the Fermat points and to the total distances. In Section 3 we provide an iterative method, converging fast to the correct Fermat points and the total distances, relevant for the multiquark potentials. We detail the cases of the baryon, the tetraquark, the pentaquark and the hexaquark. In Section 4 we conclude.

\section{Brief review of the geometrical method}

\begin{figure}[t!]
\begin{centering}
\includegraphics[width=15cm]{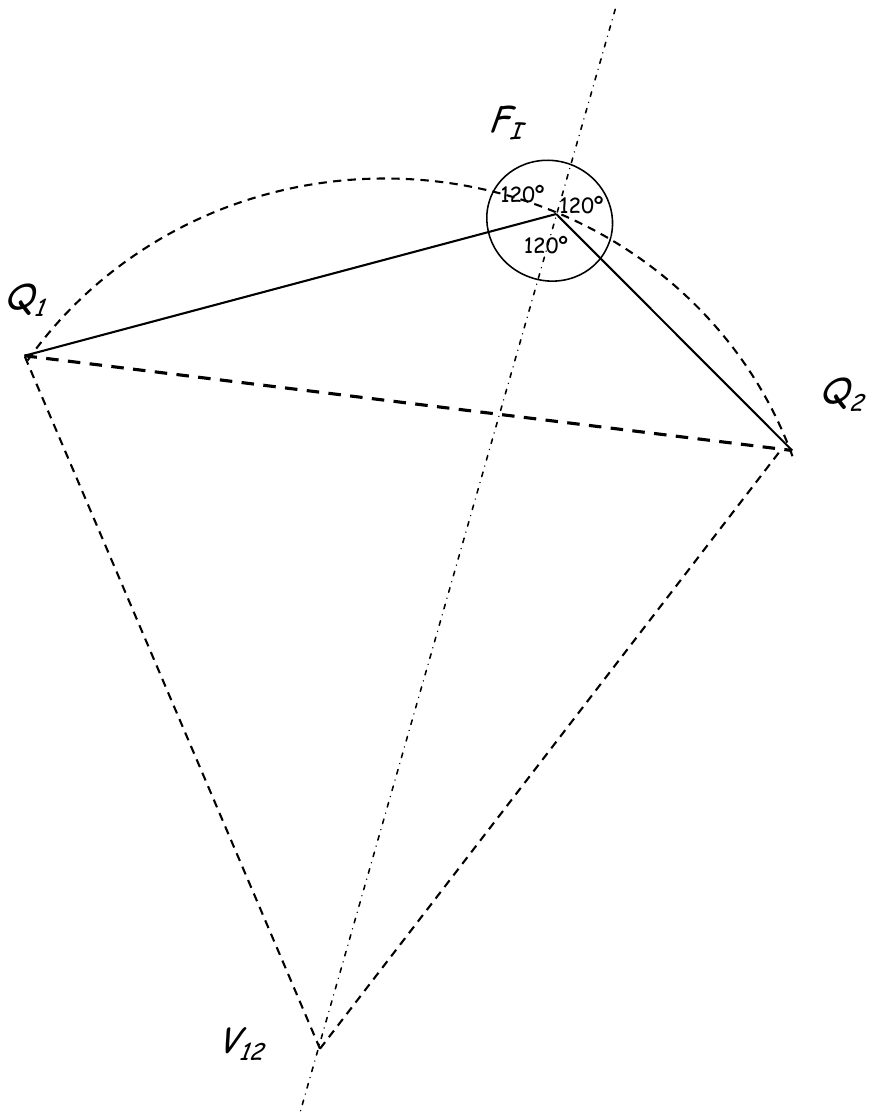}
\par\end{centering}
\caption{A step in the geometric construction of the first Fermat point of an acute triangle. Starting from the segment $Q_1 \, Q_2$, an equilateral  triangle with vertex  $V_{12}$ is constructed. The  Fermat point $F_I$ belongs to the arc of circle centered in $V_{12}$ and passing by  $Q_1$ and $ Q_2$.}
\label{equilateral}
\end{figure}

\begin{figure}[t!]
\begin{centering}
\includegraphics[width=15cm]{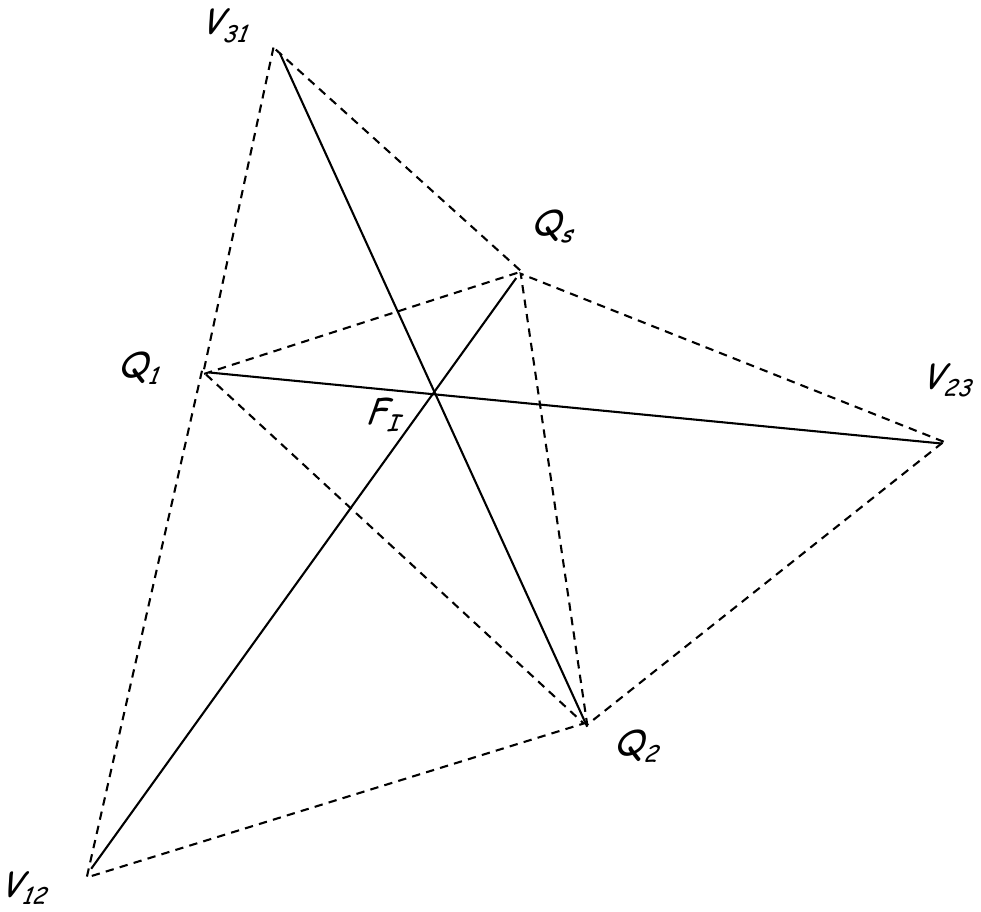}
\par\end{centering}
\caption{The geometrical method to construct the Fermat $F_I$ point of a Baryon. 
The Fermat point is at the intersection fo the three segments $V_{12}Q_3$,
 $V_{23}Q_1$ and  $V_{31}Q_2$.}
\label{baryonFermat}
\end{figure}

In an acute triangle, the Fermat point $F_I$ is the isogonic point, 
defined in eq. (\ref{isogonic}). 
To construct the isogonic point, we start by the first pair of vertices $Q_1$ and $Q_2$, noticing that the set
of points $F_I$ with fixed angle $\widehat{Q_1 \, F_I \, Q_2}=120^o$
belong to an arc of circle. Moreover this circle is centred in the other
vertex $V_{12}$ of an equilateral triangle including $Q_1$ and $Q_2$ . 
In Fig. \ref{equilateral} we show the $120^o$ arc of circle,  
the equilateral triangle, and a segment including the points  $V_{12}$ 
and $F_I$. 
Notice that the other end of this segment forms with the segments  
$Q_1 \, F_I$ and  $F_I \, Q_2$ angles of $120^o$. 
Thus the isogonic point belongs this arc of circle.
This point is at the intersection of the segments $V_{12}Q_3$,
 $V_{23}Q_1$ and  $V_{31}Q_2$. The construction
of first Fermat point $F_I$  is illustrated in Fig. \ref{baryonFermat}.
It is very simple in a geometrical perspective, and it can be done just with a compass and a rule.

\begin{figure}[t!]
\begin{centering}
\includegraphics[width=15cm]{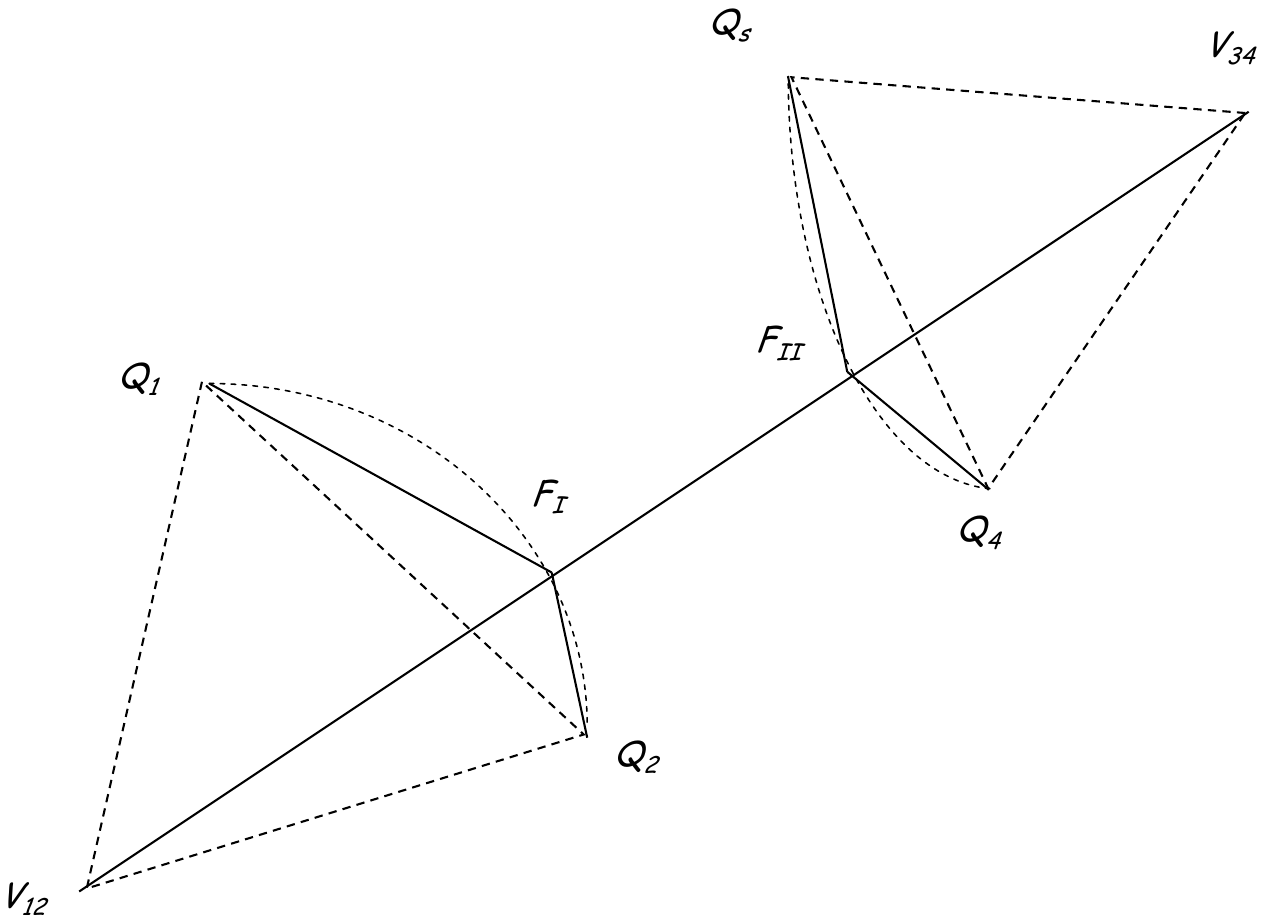}
\par\end{centering}
\caption{The geometrical method to construct the two Fermat points $F_I$ and $F_{I\hspace{-1pt}I}$ of a tetraquark. Notice that The points $Q_1$, $Q_2$, $Q_3$ and $Q_4$ are not coplanar. Thus the vertices $V_{12}$ and $V_{34}$ 
are not, from the onset determined, we only the circles where they belong. 
Nevertheless it is possible to determine tehm geometrically, knowing that the segment $V_{12}\, V_{34}$ intersects the segment $Q_1 \, Q_2$
and the segment $Q_3 \, Q_4$. Then, once the segment $V_{12}\, V_{34}$ is determined, the Fermat points $F_I$ and $F_{I\hspace{-1pt}I}$ are determined
because the distances $V_{12}\, F_I = Q_1 \, Q_2$ and 
$V_{34}\, F_{I\hspace{-1pt}I} = Q_3 \, Q_4$.}
\label{tetraquarkFermat}
\end{figure}

We now proceed with the tetraquark.
This geometrical method can be extended to construct the two Fermat points $F_I$ and $F_{I\hspace{-1pt}I}$ of a tetraquark. Notice that in the tetraquark we have four points, and thus in general the points $Q_1$, $Q_2$, $Q_3$ and $Q_4$ are not coplanar. Thus the vertices $V_{12}$ and $V_{34}$ 
are not, from the onset determined,  only the circles where they belong are
determined with the technique already used for the baryon. To determine  
the vertices, notice that the vertex $V_{12}$ must be as far as possible from the
segment $Q_3 \, Q_4$ and that the vertex $V_{34}$ must be as far as possible from the segment $Q_1 \, Q_2$. Thus we find that the segment $V_{12}\, V_{34}$ must intersect the segment $Q_1 \, Q_2$
and the segment $Q_3 \, Q_4$. Then, once the segment $V_{12}\, V_{34}$ is determined, the Fermat points $F_I$ and $F_{I\hspace{-1pt}I}$ are determined
because the distances $V_{12}\, F_I = Q_1 \, Q_2$ and 
$V_{34}\, F_{I\hspace{-1pt}I} = Q_3 \, Q_4$. This is illustrated in 
Fig. \ref{tetraquarkFermat}.

Although the solutions are simple geometrically, the algebraic
computation of the cartesian coordinates of the Fermat points $F_a$
is not simple. 
Let us consider the baryon case of triangle in a three dimensional space. 
We first need to check whether the triangle is acute.  If the triangle is acute,
we find each of the vertices  $V_{ij}$  of the equilateral triangles with a 
system of three equations, two linear equations stating that the vertex belongs to the plane of $Q_1$, $Q_2$ and $Q_3$ and to the plane of the mediatrices of 
$Q_i$ and $Q_j$ , and one quadratic equation to fix the distance of the vertex to the medium point of $Q_i$ and $Q_j$.  
We have to ensure that the vertices $V_{ij}$ point outwards the initial triangle  
$Q_1 \, Q_2 \, Q_3$. 
To find the Fermat point $F_I$ we need to get two vertices, say $V_{12}$  and $V_{23}$, and then to
intersect the segments $V_{12}Q_3$ and  $V_{23}Q_1$. So 
we first need if statements, and in the acute case we have to solve a total of 
two quadratic equations and seven linear ones.

While this system of equations and inequations is exactly solvable
for a triangle, the algebraic method gets quite difficult for larger multiquarks.
Thus another method would be welcome to compute the Fermat points 
$F_a$ and the total distance 
$d= \sum _{i,a} r_{i\,a}$.

\section{Iterative method}

We now propose an iterative method, designed for a computational 
determination of the confining potential of multiquarks.

\begin{figure}[t!]
\begin{centering}
\includegraphics[width=7.5cm]{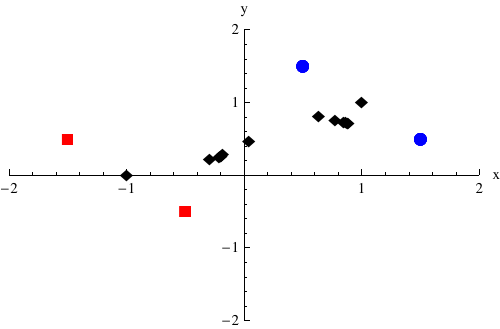}
\includegraphics[width=7.5cm]{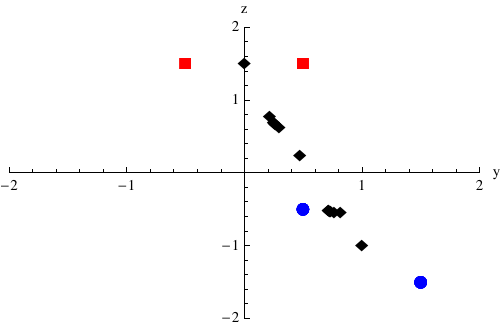}
\includegraphics[width=7.5cm]{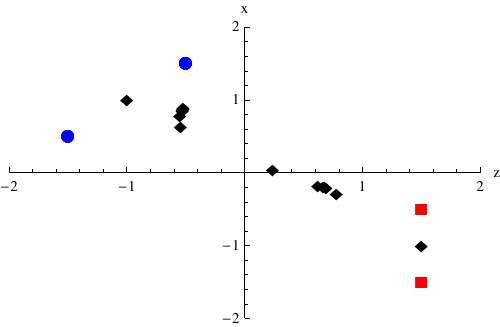}
\par\end{centering}
\caption{Convergence of the numerical iterative method
method to construct the two Fermat points $F_I$ and $F_{I\hspace{-1pt}I}$ of an arbitrary tetraquark. The quarks are depicted as circles, the antiquarks as squares, and the converging Fermat Points as diamonds. We show projections in the $xy$, $yz$ and $zx$ planes. Visually, after 6 iterations the iteration has converged.  On average, a $10^{-6}$ precision is achieved for the tetraquark after 20 iterations.}
\label{convfig}
\end{figure}

We start from the triquark case of the baryon. We use the notation,
\bea
\mbf r_i &=&(x_i,y_i,z_i)\ ,
\nonumber \\
\mbf r_I &=& (x_I ,y_I ,z_I )\ ,
\nonumber \\
r_{i\,I} &=& \sqrt{  (x_I -x_i)^2+ (y_I -y_i)^2+ (z_I -z_i)^2 }\ .
\eea
Minimizing the total distance $r_{1\,I}+r_{2\,I}+r_{3\,I}$ with regards to
the three coordinates of the Fermat point $F_I$, we get the
system equations,
\bea
x_I  &=& { {x_1 \over r_{1\,I} }+ {x_2 \over r_{2\,I} }+{x_3 \over r_{3\,I} }\over
{1 \over r_{1\,I} }+ {1 \over r_{2\,I} }+{1 \over r_{3\,I} }} \ ,
\nonumber \\
y_I  &=& { {y_1 \over r_{1\,I} }+ {y_2 \over r_{2\,I} }+{y_3 \over r_{3\,I} }\over
{1 \over r_{1\,I} }+ {1 \over r_{2\,I} }+{1 \over r_{3\,I} }} \ ,
\nonumber \\
z_I  &=& { {z_1 \over r_{1\,I} }+ {z_2 \over r_{2\,I} }+{z_3 \over r_{3\,I} }\over
{1 \over r_{1\,I} }+ {1 \over r_{2\,I} }+{1 \over r_{3\,I} }} \ .
\label{nlin Fermat}
\eea
which is non-linear. To solve these algebraic equations, of the form
\bea
 x =X(x ,y ,z )\ , 
\nonumber \\
y  = Y(x ,y ,z )\ , 
\nonumber\\
z  = Z(x ,y ,z )\ ,
\eea 
an iterative method can be used. With a relaxation coefficient $\omega$, 
we iterate the series,
\bea
x_{n+1} &=&  \omega X(x_n, y_n,z_n) + (1-\omega) x_n \ ,
\nonumber \\
y_{n+1} &=&  \omega Y(x_n, y_n,z_n) + (1-\omega) y_n  \ ,
\nonumber \\
z_{n+1} &=&  \omega Z(x_n, y_n,z_n)  + (1-\omega) z_n \ ,
\label{iterat Fermat}
\eea
starting with, as an initial guess, the barycentre  
\be
{\mbf r_I}_0= {\mbf r_1 + \mbf r_2 + \mbf r_3 \over 3}\ .
\ee

A first numerical check shows that the method converges 
rapidly to the Fermat path of the  triangle. We get get results 
accurate to a precision of $10^{-6}$, for the total distance
$d=r_{1\,I}+r_{2\,I}+r_{3\,I}$ after a number
of iterations of the order of 12, for an optimized relaxation 
factor $\omega=1.7$.

Thus we extend our iterarive method to the cases of the
next multiquarks. 
We simply get a system of there equations per Fermat point. 
For the tetraquark the equations are,
\bea
x_I  &=& { {x_1 \over r_{1\,I} }+ {x_2 \over r_{2\,I} }+{x_{I\hspace{-1pt}I} \over r_{{I\hspace{-1pt}I}\,I} }\over
{1 \over r_{1\,I} }+ {1 \over r_{2\,I} }+{1 \over r_{{I\hspace{-1pt}I}\,I} }} \ ,
\nonumber \\
y_I  &=& { {y_1 \over r_{1\,I} }+ {y_2 \over r_{2\,I} }+{y_{I\hspace{-1pt}I} \over r_{{I\hspace{-1pt}I}\,I} }\over
{1 \over r_{1\,I} }+ {1 \over r_{2\,I} }+{1 \over r_{{I\hspace{-1pt}I}\,I} }} \ ,
\nonumber \\
z_I  &=& { {z_1 \over r_{1\,I} }+ {z_2 \over r_{2\,I} }+{z_{I\hspace{-1pt}I} \over r_{{I\hspace{-1pt}I}\,I} }\over
{1 \over r_{1\,I} }+ {1 \over r_{2\,I} }+{1 \over r_{{I\hspace{-1pt}I}\,I} }} \ ,
\nonumber \\
x_{I\hspace{-1pt}I}  &=& { {x_3 \over r_{3\,{I\hspace{-1pt}I}} }+ {x_4 \over r_{4\,{I\hspace{-1pt}I}} }+{x_I \over r_{I\,{I\hspace{-1pt}I}} }\over
{1 \over r_{3\,{I\hspace{-1pt}I}} }+ {1 \over r_{4\,{I\hspace{-1pt}I}} }+{1 \over r_{I\,{I\hspace{-1pt}I}} }} \ ,
\nonumber \\
y_{I\hspace{-1pt}I}  &=& { {y_3 \over r_{3\,{I\hspace{-1pt}I}} }+ {y_4 \over r_{4\,{I\hspace{-1pt}I}} }+{y_I \over r_{I\,{I\hspace{-1pt}I}} }\over
{1 \over r_{3\,{I\hspace{-1pt}I}} }+ {1 \over r_{4\,{I\hspace{-1pt}I}} }+{1 \over r_{I\,{I\hspace{-1pt}I}} }} \ ,
\nonumber \\
z_{I\hspace{-1pt}I}  &=& { {z_3 \over r_{3\,{I\hspace{-1pt}I}} }+ {z_4 \over r_{4\,{I\hspace{-1pt}I}} }+{z_I \over r_{I\,{I\hspace{-1pt}I}} }\over
{1 \over r_{3\,{I\hspace{-1pt}I}} }+ {1 \over r_{4\,{I\hspace{-1pt}I}} }+{1 \over r_{I\,{I\hspace{-1pt}I}} }} \ ,
\label{nlin tetra}
\eea
and we also check that the iterations converge fast to the Fermat paths
for the tetraquark. 
We illustrate graphically, for an arbitrary geometric configuration,  
the convergence in the case of a tetraquark in Fig. \ref{convfig}.
For the pentaquark the equations are,
\bea
x_I  &=& { {x_1 \over r_{1\,I} }+ {x_2 \over r_{2\,I} }+{x_{I\hspace{-1pt}I\hspace{-1pt}I} \over r_{{I\hspace{-1pt}I\hspace{-1pt}I}\,I} }\over
{1 \over r_{1\,I} }+ {1 \over r_{2\,I} }+{1 \over r_{{I\hspace{-1pt}I\hspace{-1pt}I}\,I} }} \ ,
\nonumber \\
y_I  &=& { {y_1 \over r_{1\,I} }+ {y_2 \over r_{2\,I} }+{y_{I\hspace{-1pt}I\hspace{-1pt}I} \over r_{{I\hspace{-1pt}I\hspace{-1pt}I}\,I} }\over
{1 \over r_{1\,I} }+ {1 \over r_{2\,I} }+{1 \over r_{{I\hspace{-1pt}I\hspace{-1pt}I}\,I} }} \ ,
\nonumber \\
z_I  &=& { {z_1 \over r_{1\,I} }+ {z_2 \over r_{2\,I} }+{z_{I\hspace{-1pt}I\hspace{-1pt}I} \over r_{{I\hspace{-1pt}I\hspace{-1pt}I}\,I} }\over
{1 \over r_{1\,I} }+ {1 \over r_{2\,I} }+{1 \over r_{{I\hspace{-1pt}I\hspace{-1pt}I}\,I} }} \ ,
\nonumber \\
x_{I\hspace{-1pt}I}  &=& { {x_3 \over r_{3\,{I\hspace{-1pt}I}} }+ {x_5 \over r_{5\,{I\hspace{-1pt}I}} }+{x_{I\hspace{-1pt}I\hspace{-1pt}I} \over r_{{I\hspace{-1pt}I\hspace{-1pt}I}\,{I\hspace{-1pt}I}} }\over
{1 \over r_{3\,{I\hspace{-1pt}I}} }+ {1 \over r_{4\,{I\hspace{-1pt}I}} }+{1 \over r_{{I\hspace{-1pt}I\hspace{-1pt}I}\,{I\hspace{-1pt}I}} }} \ ,
\nonumber \\
y_{I\hspace{-1pt}I}  &=& { {y_3 \over r_{3\,{I\hspace{-1pt}I}} }+ {y_4 \over r_{4\,{I\hspace{-1pt}I}} }+{y_{I\hspace{-1pt}I\hspace{-1pt}I} \over r_{{I\hspace{-1pt}I\hspace{-1pt}I}\,{I\hspace{-1pt}I}} }\over
{1 \over r_{3\,{I\hspace{-1pt}I}} }+ {1 \over r_{4\,{I\hspace{-1pt}I}} }+{1 \over r_{{I\hspace{-1pt}I\hspace{-1pt}I}\,{I\hspace{-1pt}I}} }} \ ,
\nonumber \\
z_{I\hspace{-1pt}I}  &=& { {z_3 \over r_{3\,{I\hspace{-1pt}I}} }+ {z_4 \over r_{4\,{I\hspace{-1pt}I}} }+{z_{I\hspace{-1pt}I\hspace{-1pt}I} \over r_{{I\hspace{-1pt}I\hspace{-1pt}I}\,{I\hspace{-1pt}I}} }\over
{1 \over r_{3\,{I\hspace{-1pt}I}} }+ {1 \over r_{4\,{I\hspace{-1pt}I}} }+{1 \over r_{{I\hspace{-1pt}I\hspace{-1pt}I}\,{I\hspace{-1pt}I}} }} \ ,
\nonumber \\
x_{I\hspace{-1pt}I\hspace{-1pt}I}  &=& { {x_I \over r_{I\,{I\hspace{-1pt}I\hspace{-1pt}I}} }+ {x_{I\hspace{-1pt}I} \over r_{{I\hspace{-1pt}I}\,{I\hspace{-1pt}I\hspace{-1pt}I}} }+{x_4 \over r_{4\,{I\hspace{-1pt}I\hspace{-1pt}I}} }\over
{1 \over r_{I\,{I\hspace{-1pt}I\hspace{-1pt}I}} }+ {1 \over r_{{I\hspace{-1pt}I}\,{I\hspace{-1pt}I\hspace{-1pt}I}} }+{1 \over r_{5\,{I\hspace{-1pt}I\hspace{-1pt}I}} }} \ ,
\nonumber \\
y_{I\hspace{-1pt}I\hspace{-1pt}I}  &=& { {y_I \over r_{I\,{I\hspace{-1pt}I\hspace{-1pt}I}} }+ {y_{I\hspace{-1pt}I} \over r_{{I\hspace{-1pt}I}\,{I\hspace{-1pt}I\hspace{-1pt}I}} }+{y_5 \over r_{5\,{I\hspace{-1pt}I\hspace{-1pt}I}} }\over
{1 \over r_{I\,{I\hspace{-1pt}I\hspace{-1pt}I}} }+ {1 \over r_{{I\hspace{-1pt}I}\,{I\hspace{-1pt}I\hspace{-1pt}I}} }+{1 \over r_{5\,{I\hspace{-1pt}I\hspace{-1pt}I}} }} \ ,
\nonumber \\
z_{I\hspace{-1pt}I\hspace{-1pt}I}  &=& { {z_I \over r_{I\,{I\hspace{-1pt}I\hspace{-1pt}I}} }+ {z_{I\hspace{-1pt}I} \over r_{{I\hspace{-1pt}I}\,{I\hspace{-1pt}I\hspace{-1pt}I}} }+{z_5 \over r_{5\,{I\hspace{-1pt}I\hspace{-1pt}I}} }\over
{1 \over r_{I\,{I\hspace{-1pt}I\hspace{-1pt}I}} }+ {1 \over r_{{I\hspace{-1pt}I}\,{I\hspace{-1pt}I\hspace{-1pt}I}} }+{1 \over r_{5\,{I\hspace{-1pt}I\hspace{-1pt}I}} }} \ ,
\label{nlin penta}
\eea
\begin{table}[t]
\begin{centering}
\begin{tabular}{c|c|c|cccccc}
\hline
\textbf{multiquark  }& \textbf{number  }  & \textbf{optimal}  & 
\multicolumn{6}{c}{\textbf{average number of iterations for a }$p=$}
\tabularnewline
& \textbf{of quarks }  & $\omega$ & 
$10^{-1}$ &$10^{-2}$ &$10^{-3}$ &
$10^{-4}$ &$10^{-5}$ &$10^{-6}$ 
\tabularnewline
\hline
meson & 2 & - & 1 & 
1 & 1 & 1 & 1 & 1 
\tabularnewline
baryon &  3 & 1.7 & 
1.1 & 1.9 & 3 & 5 & 8 & 12 
\tabularnewline
tetraquark & 4 & 1.4 & 
1.7 & 2.7 & 5 & 9 & 14 & 20  
\tabularnewline
pentaquark & 5 & 1.5 & 
1.5 & 2.5 & 5 & 10 & 18 & 29
\tabularnewline
hexaquark & 6 & 1.4 & 
1.5 & 2.8 & 5 & 10 & 16 & 26
\tabularnewline
\hline
\end{tabular}
\par\end{centering}
\caption{Convergence of the iterative method for different multiquarks, based on a million of random generated quark
positions, for each multiquark, and for different
precisions $p= {\Delta d \over d}$, where $d$ is the total Fermat distance. The values of the relaxation factor $\omega$ are the ones which minimize the convergence time up to a precision $p=10^{-6}$. }
\label{convtable}
\end{table}
for the hexaquark the equations are,
\bea
x_I  &=& { {x_1 \over r_{1\,I} }+ {x_2 \over r_{2\,I} }+{x_{I\hspace{-1pt}V} \over r_{{I\hspace{-1pt}V}\,I} }\over
{1 \over r_{1\,I} }+ {1 \over r_{2\,I} }+{1 \over r_{{I\hspace{-1pt}V}\,I} }} \ ,
\nonumber \\
y_I  &=& { {y_1 \over r_{1\,I} }+ {y_2 \over r_{2\,I} }+{y_{I\hspace{-1pt}V} \over r_{{I\hspace{-1pt}V}\,I} }\over
{1 \over r_{1\,I} }+ {1 \over r_{2\,I} }+{1 \over r_{{I\hspace{-1pt}V}\,I} }} \ ,
\nonumber \\
z_I  &=& { {z_1 \over r_{1\,I} }+ {z_2 \over r_{2\,I} }+{z_{I\hspace{-1pt}V} \over r_{{I\hspace{-1pt}V}\,I} }\over
{1 \over r_{1\,I} }+ {1 \over r_{2\,I} }+{1 \over r_{{I\hspace{-1pt}V}\,I} }} \ ,
\nonumber \\
x_{I\hspace{-1pt}I}  &=& { {x_3 \over r_{3\,{I\hspace{-1pt}I}} }+ {x_4 \over r_{4\,{I\hspace{-1pt}I}} }+{x_{I\hspace{-1pt}V} \over r_{{I\hspace{-1pt}V}\,{I\hspace{-1pt}I}} }\over
{1 \over r_{3\,{I\hspace{-1pt}I}} }+ {1 \over r_{4\,{I\hspace{-1pt}I}} }+{1 \over r_{{I\hspace{-1pt}V}\,{I\hspace{-1pt}I}} }} \ ,
\nonumber \\
y_{I\hspace{-1pt}I}  &=& { {y_3 \over r_{3\,{I\hspace{-1pt}I}} }+ {y_4 \over r_{4\,{I\hspace{-1pt}I}} }+{y_{I\hspace{-1pt}V} \over r_{{I\hspace{-1pt}V}\,{I\hspace{-1pt}I}} }\over
{1 \over r_{3\,{I\hspace{-1pt}I}} }+ {1 \over r_{4\,{I\hspace{-1pt}I}} }+{1 \over r_{{I\hspace{-1pt}V}\,{I\hspace{-1pt}I}} }} \ ,
\nonumber \\
z_{I\hspace{-1pt}I}  &=& { {z_3 \over r_{3\,{I\hspace{-1pt}I}} }+ {z_4 \over r_{4\,{I\hspace{-1pt}I}} }+{z_{I\hspace{-1pt}V} \over r_{{I\hspace{-1pt}V}\,{I\hspace{-1pt}I}} }\over
{1 \over r_{3\,{I\hspace{-1pt}I}} }+ {1 \over r_{4\,{I\hspace{-1pt}I}} }+{1 \over r_{{I\hspace{-1pt}V}\,{I\hspace{-1pt}I}} }} \ ,
\nonumber \\
x_{I\hspace{-1pt}I\hspace{-1pt}I}  &=& { {x_5 \over r_{5\,{I\hspace{-1pt}I\hspace{-1pt}I}} }+ {x_6 \over r_{6\,{I\hspace{-1pt}I\hspace{-1pt}I}} }+{x_{I\hspace{-1pt}V} \over r_{{I\hspace{-1pt}V}\,{I\hspace{-1pt}I\hspace{-1pt}I}} }\over
{1 \over r_{5\,{I\hspace{-1pt}I\hspace{-1pt}I}} }+ {1 \over r_{6\,{I\hspace{-1pt}I\hspace{-1pt}I}} }+{1 \over r_{{I\hspace{-1pt}V}\,{I\hspace{-1pt}I\hspace{-1pt}I}} }} \ ,
\nonumber \\
y_{I\hspace{-1pt}I\hspace{-1pt}I}  &=& { {y_5 \over r_{5\,{I\hspace{-1pt}I\hspace{-1pt}I}} }+ {y_6 \over r_{6\,{I\hspace{-1pt}I\hspace{-1pt}I}} }+{y_{I\hspace{-1pt}V} \over r_{{I\hspace{-1pt}V}\,{I\hspace{-1pt}I\hspace{-1pt}I}} }\over
{1 \over r_{5\,{I\hspace{-1pt}I\hspace{-1pt}I}} }+ {1 \over r_{6\,{I\hspace{-1pt}I\hspace{-1pt}I}} }+{1 \over r_{{I\hspace{-1pt}V}\,{I\hspace{-1pt}I\hspace{-1pt}I}} }} \ ,
\nonumber \\
z_{I\hspace{-1pt}I\hspace{-1pt}I}  &=& { {z_5 \over r_{5\,{I\hspace{-1pt}I\hspace{-1pt}I}} }+ {z_6 \over r_{6\,{I\hspace{-1pt}I\hspace{-1pt}I}} }+{z_{I\hspace{-1pt}V} \over r_{{I\hspace{-1pt}V}\,{I\hspace{-1pt}I\hspace{-1pt}I}} }\over
{1 \over r_{5\,{I\hspace{-1pt}I\hspace{-1pt}I}} }+ {1 \over r_{6\,{I\hspace{-1pt}I\hspace{-1pt}I}} }+{1 \over r_{{I\hspace{-1pt}V}\,{I\hspace{-1pt}I\hspace{-1pt}I}} }} \ ,
\nonumber \\
x_{I\hspace{-1pt}V}  &=& { {x_I \over r_{I\,{I\hspace{-1pt}V}} }+ {x_{I\hspace{-1pt}I} \over r_{{I\hspace{-1pt}I}\,{I\hspace{-1pt}V}} }+{x_{I\hspace{-1pt}I\hspace{-1pt}I} \over r_{{I\hspace{-1pt}I\hspace{-1pt}I}\,{I\hspace{-1pt}V}} }\over
{1 \over r_{I\,{I\hspace{-1pt}V}} }+ {1 \over r_{{I\hspace{-1pt}I}\,{I\hspace{-1pt}V}} }+{1 \over r_{{I\hspace{-1pt}I\hspace{-1pt}I}\,{I\hspace{-1pt}V}} }} \ ,
\nonumber \\
y_{I\hspace{-1pt}V}  &=& { {y_I \over r_{I\,{I\hspace{-1pt}V}} }+ {y_{I\hspace{-1pt}I} \over r_{{I\hspace{-1pt}I}\,{I\hspace{-1pt}V}} }+{y_{I\hspace{-1pt}I\hspace{-1pt}I} \over r_{{I\hspace{-1pt}I\hspace{-1pt}I}\,{I\hspace{-1pt}V}} }\over
{1 \over r_{I\,{I\hspace{-1pt}V}} }+ {1 \over r_{{I\hspace{-1pt}I}\,{I\hspace{-1pt}V}} }+{1 \over r_{{I\hspace{-1pt}I\hspace{-1pt}I}\,{I\hspace{-1pt}V}} }} \ ,
\nonumber \\
z_{I\hspace{-1pt}V}  &=& { {z_I \over r_{I\,{I\hspace{-1pt}V}} }+ {z_{I\hspace{-1pt}I} \over r_{{I\hspace{-1pt}I}\,{I\hspace{-1pt}V}} }+{z_{I\hspace{-1pt}I\hspace{-1pt}I} \over r_{{I\hspace{-1pt}I\hspace{-1pt}I}\,{I\hspace{-1pt}V}} }\over
{1 \over r_{I\,{I\hspace{-1pt}V}} }+ {1 \over r_{{I\hspace{-1pt}I}\,{I\hspace{-1pt}V}} }+{1 \over r_{{I\hspace{-1pt}I\hspace{-1pt}I}\,{I\hspace{-1pt}V}} }} \ ,
\label{nlin hexa}
\eea
and for larger multiquarks the extension of these equations is obvious.

\begin{figure}[t!]
\begin{centering}
\includegraphics[width=15cm]{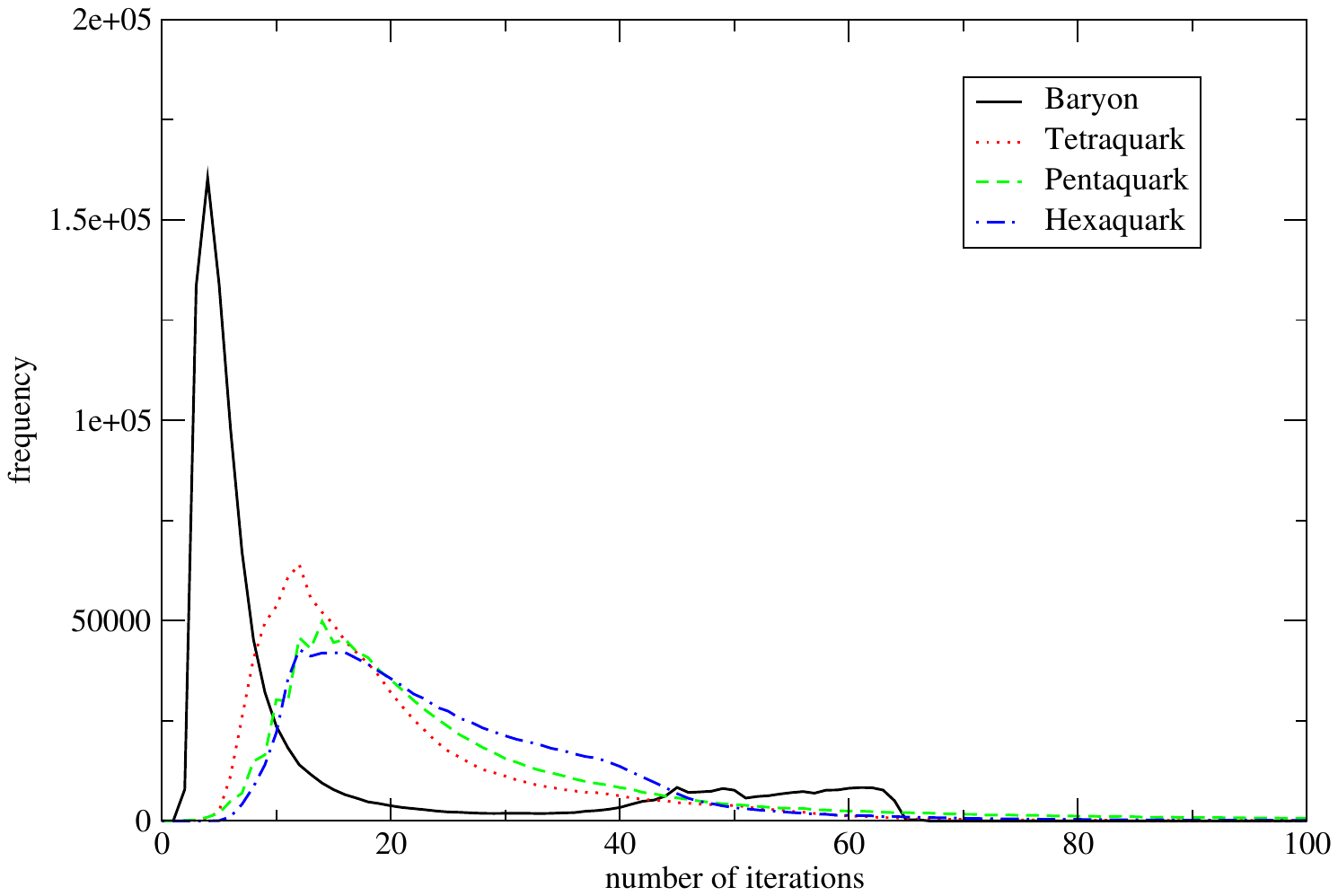}
\par\end{centering}
\caption{ Distribution of the number of iterations needed to converge to a 
precision $p=10^{-6}$, based on a million
of random generated quark positions, for each multiquark.
 }
\label{iters}
\end{figure}

The convergence of the method for the first multiquark systems is shown in Table \ref{convtable} and in  the Fig. \ref{iters}

We study numerically the convergence of the method, with a 
randomly generated sample of $10^6$ geometric configurations,
for each of the multiquarks. We first optimize the relaxation factor 
$\omega$ in order to reduce the needed number of iteration steps, 
to get results accurate to a precision of $10^{-6}$, 
for the total distance $d= \sum _{i,a} r_{i\,a}$.
Since the number of iterations is not a constant, we 
demand the $\omega$ minimizing the average $\bar d$
over the sample of geometric configurations. The optimized
$\omega$ and the resulting $\bar d$ are shown in Table \ref{convtable}
for the baryon, the tetraquark, the pentaquark and the hexaquark.
Notice that the convergence is fast, even for the hexaquark.

Then we compute the distribution of the number of convergences
we have per number of iterations. The number of configurations
is large, and so we don't need to join different numbers of iterations in 
a bin. In  Fig. \ref{iters} we show the distribution of the number of 
distribution of the number of convergences
we have per number of iterations for the baryon, the tetraquark, 
the pentaquark and the hexaquark.

\section{Conclusion}

We study an iterative method to find the Fermat points and Fermat 
distances in multiquarks. This method replaces the geometrical
method, which has only been applied so far to the baryon (triquark).

The method is very simple to programme, and it converges both in the 
case of acute angles(smaller than $120^o$)  and of larger angles.
This avoids the problem of checking for non-acute angle, simple
for a Baryon, but harder for the larger multiquarks.

This method is suited to be used in the solution of the Schr\"odinger 
equation in a quark model, where we have to compute the 
potential for several different positions of the quarks and antiquarks.

Thus we use, as a convergence criterion, the precision of
$10^{-6}$, one part per million, in the total distance of
the Fermat paths defined in Fig. \ref{first five}.
Even for this extremely fine precision, far beyond the normal 
precision of quark models,  the method is quite fast,
as show in Table \ref{convtable}.
The number of necessary iterations $n$ to achieve
a precision $p= {\Delta d  \over d}$, grows 
quadratically with the number of desired correct decimal 
cases, $- \log p$ is proportional to $n^2$.

The present computational technique enables the precision 
studies of the multiquarks with string confinement in the quark model.

\section*{Acknowledgments}
This study was possible due to the FCT grants PDCT/FP/63923/2005 and POCI/FP/81933/2007.

\bibliographystyle{elsart-num}

\end{document}